%% file: paper.tex
\documentclass[twocolumn,showpacs,aps,prl,superscriptaddress]{revtex4}
\usepackage{graphicx}
\usepackage{dcolumn}
\usepackage{amsmath}
\usepackage{epsfig}

\RequirePackage{xspace}

\usepackage{relsize}
\def\babar{\mbox{\slshape B\kern-0.1em{\smaller A}\kern-0.1em
    B\kern-0.1em{\smaller A\kern-0.2em R}}}
\def\piz   {\ensuremath{\pi^0}\xspace}
\def\pip   {\ensuremath{\pi^+}\xspace}
\def\pim   {\ensuremath{\pi^-}\xspace}
\def\Kbar  {\kern 0.2em\overline{\kern -0.2em K}{}\xspace}

\def\Kp    {\ensuremath{K^+}\xspace}
\def\Km    {\ensuremath{K^-}\xspace}

\def\KS    {\ensuremath{K^0_{\scriptscriptstyle S}}\xspace} 
 
\def\Dbar    {\kern 0.2em\overline{\kern -0.2em D}{}\xspace}

\def\Dz      {\ensuremath{D^0}\xspace}
\def\Dzb     {\ensuremath{\Dbar^0}\xspace}
\def\DzDzb   {\ensuremath{\Dz {\kern -0.16em \Dzb}}\xspace}
\def\Bbar    {\kern 0.18em\overline{\kern -0.18em B}{}\xspace}
\def\BB      {\ensuremath{B\Bbar}\xspace} 
\mathchardef\Upsilon="7107
\def\Y#1S{\ensuremath{\Upsilon{(#1S)}}\xspace}
\def\FourS {\Y4S}
\def\BR         {{\ensuremath{\cal B}\xspace}}
\def\mes        {\mbox{$m_{\rm ES}$}\xspace}
\def\DeltaE     {\mbox{$\Delta E$}\xspace}
\newcommand{\gev}{\ensuremath{\mathrm{\,Ge\kern -0.1em V}}\xspace}
\newcommand{\mev}{\ensuremath{\mathrm{\,Me\kern -0.1em V}}\xspace}
\newcommand{\gevc}{\ensuremath{{\mathrm{\,Ge\kern -0.1em V\!/}c}}\xspace}
\newcommand{\mevc}{\ensuremath{{\mathrm{\,Me\kern -0.1em V\!/}c}}\xspace}
\newcommand{\gevcc}{\ensuremath{{\mathrm{\,Ge\kern -0.1em V\!/}c^2}}\xspace}
\newcommand{\mevcc}{\ensuremath{{\mathrm{\,Me\kern -0.1em V\!/}c^2}}\xspace}
\def\invfb   {\ensuremath{\mbox{\,fb}^{-1}}\xspace}

\def\ra                 {\ensuremath{\rightarrow}\xspace}
\def\to                 {\ensuremath{\rightarrow}\xspace}
\newcommand{\stat}{\ensuremath{\mathrm{(stat)}}\xspace}
\newcommand{\syst}{\ensuremath{\mathrm{(syst)}}\xspace}
\def\pep2{PEP-II}

\def\CP                {\ensuremath{C\!P}\xspace}
\def\CPp                {\ensuremath{C\!P\!+}\xspace}
\def\CPm                {\ensuremath{C\!P\!-}\xspace}

\newcommand{\jprlBase}       {Phys.\ Rev.\ Lett.\xspace}
\newcommand{\jprBase}        {Phys.\ Rev.\xspace}
\newcommand{\jplBase}        {Phys.\ Lett.\xspace}

\newcommand{\nimBaseC}       {Nucl.\ Instr.\ and Methods\xspace}

\newcommand{\nim}       [1]  {\nimBaseC~{\bf #1}}

\newcommand{\jpl}       [1]  {\jplBase\ {\bf #1}}

\newcommand{\jprl}      [1]  {\jprlBase\ {\bf #1}}
\newcommand{\pr}        [1]  {\jprBase\ {\bf #1}}

\newcommand{\BABARPubYear}    {05}
\newcommand{\BABARPubNumber} {051}

\newcommand{\SLACPubNumber}{11610}

\def\figurebox#1#2#3{%
    \def\arg{#3}%
    \ifx\arg\empty
    {\hfill\vbox{\hsize#2\hrule\hbox to #2{\vrule\hfill\vbox to #1{\hsize#2\vfill}\vrule}\hrule}\hfill}%
    \else
    {\hfill\epsfbox{#3}\hfill}%
    \fi}

\def\btodk   {\ensuremath {B^-{\to}D^0 K^-}}

\def\btodp   {\ensuremath {B^-{\to}D^0 \pi^-}}

\def\btodh   {\ensuremath {B^-{\to}D^0 h^-}}

\def\dotopp  {\ensuremath {D^0{\to}\pi^-\pi^+}}

\begin{document}

\noindent
\babar-PUB-\BABARPubYear/\BABARPubNumber\\
SLAC-PUB-\SLACPubNumber
\vskip 0.4cm

\title{
  {\large \bf
    Measurements of the branching fractions and \boldmath{\CP}-asymmetries of
    \boldmath{$B \to D^0_{\CP} K$} decays
  }
}

\input authors_nov2005

\date{\today}

\begin{abstract}
We present a study of the decay $B^-{\to}D^0_{(\CP)}K^-$ and its charge
conjugate, where $D^0_{(\CP)}$ is 
reconstructed in \CP-even, \CP-odd, and non-\CP flavor eigenstates,
based on a sample of 232 million $\Upsilon(4S){\to}\BB$ decays
collected with the \babar\ detector at the PEP-II $e^+e^-$ storage ring.
We measure the partial-rate charge asymmetries $A_{\CP\pm}$ and the
ratios $R_{\CP\pm}$ of the $B{\to}D^0K$ decay  
branching fractions as measured in $\CP\pm$ and non-$\CP$ $D^0$ decays:
$A_{\CPp} = 0.35\pm 0.13\stat\pm 0.04\syst$,
$A_{\CPm} =-0.06\pm 0.13\stat\pm 0.04\syst$,
$R_{\CP+} = 0.90\pm 0.12\stat\pm 0.04\syst$,
$R_{\CP-} = 0.86\pm 0.10\stat\pm 0.05\syst$.
\end{abstract}

\pacs{11.30.Er,13.25.Hw,14.40.Nd}

\maketitle

A stringent test of the flavor and \CP\ sector of the Standard Model
can be obtained from the measurements, in $B$ meson decays, of the
sides and angles of the unitarity triangle, which are related to the
elements of the Cabibbo-Kobayashi-Maskawa matrix $V$.
A theoretically clean measurement of the
angle $\gamma=\arg(-V_{ud}V_{ub}^*/V_{cd}V_{cb}^*)$ can  
be obtained from the study of $B^-{\to}D^{(*)0}K^{(*)-}$
decays~\cite{chargeconj} 
by exploiting the interference between the $b\ra c\bar{u}s$ and
$b\ra u\bar{c}s$ decay amplitudes~\cite{gronau1991,others_btdk}.
Among the proposed methods, the one originally suggested by Gronau,
London, and Wyler (GLW) exploits the interference between $B^-{\to} D^0K^-$ and
$B^-{\to} \Dzb K^-$ when the $D^0$ and $\Dzb$ mesons decay to the same
\CP eigenstate.

The results of the GLW analyses are usually expressed in terms
of the ratios $R_{\CP\pm}$ of charge-averaged partial rates and of the
partial-rate charge asymmetries $A_{\CP\pm}$,  
\begin{eqnarray}
  &&R_{\CP\pm} = \frac{\Gamma(B^-{\to}\Dz_{\CP\pm}K^-) +
    \Gamma(B^+{\to}\Dz_{\CP\pm}K^+)} {\left[\Gamma(B^-{\to}\Dz K^-)+\Gamma(B^+{\to}\Dzb K^+)\right]/2}\,,\ \ \ \  \\ 
&&A_{\CP\pm}=\frac{\Gamma(B^-{\to}\Dz_{\CP\pm}K^-)-\Gamma(B^+{\to}\Dz_{\CP\pm}K^+)}{\Gamma(B^-{\to}\Dz_{\CP\pm}K^-)+\Gamma(B^+{\to}\Dz_{\CP\pm}K^+)}\,.\ \ \ \
\end{eqnarray}
Here, $D^0_{\CP\pm} = (\Dz \pm \Dzb)/\sqrt{2}$ are the 
\CP eigenstates of the neutral $D$ meson system, and we have followed
the notation used in~\cite{gronau1998}.
Neglecting $D^0{-}\Dzb$ mixing~\cite{dmixing},
the observables 
$R_{\CP\pm}$ and $A_{\CP\pm}$ are related to the angle $\gamma$, the 
magnitude $r$ of the ratio of the amplitudes for the
processes $B^-{\to} \Dzb K^-$ and $B^-{\to} D^0 K^-$, and the relative
strong phase  $\delta$ between these two amplitudes, through the
relations $R_{\CP\pm}=1+r^2\pm 2r\cos\delta\cos\gamma$ and 
$A_{\CP\pm}=\pm 2r\sin\delta\sin\gamma/R_{\CP\pm}$~\cite{gronau1991}. 
Theoretical expectations for $r$ are in the range $\approx
0.1-0.2$~\cite{gronau1991,gronau2003}, in agreement with the 90\% C.L. upper
limits on $r$ set by \babar\ ($r<0.23$) and Belle ($r<0.18$) through the study
of $B^-{\to} DK^-, D{\to} K^+\pi^-$ decays~\cite{babar_ads}. 

In this paper we present the measurements of $R_{\CP\pm}$ and
$A_{\CP\pm}$.
The ratios $R_{\CP\pm}$ are computed using
the relations $R_{\CP\pm}\simeq R_{\pm}/R$, where the quantities
$R$ and $R_{\pm}$ are defined as:
\begin{eqnarray}
&& R\ = \frac{\BR(B^-{\to}\Dz K^-)+\BR(B^+{\to}\Dzb
  K^+)}{\BR(B^-{\to}\Dz\pi^-)+\BR(B^+{\to}\Dzb\pi^+)}\,,\label{eq:r}\\
&& R_{\pm}
  =\frac{\BR(B^-{\to}\Dz_{\CP\pm}K^-)+\BR(B^+{\to}\Dz_{\CP\pm}K^+)}{\BR(B^-{\to}\Dz_{\CP\pm}\pi^-)+\BR(B^+{\to}\Dz_{\CP\pm}\pi^+)}\,.\ \ \label{eq:rpm}
\end{eqnarray}
Several systematic uncertainties cancel out in the measurement of these
double ratios.
We also express the $\CP$-sensitive observables
in terms of three independent quantities:
\begin{eqnarray}
&&x_\pm=\frac{R_{\CP+}(1\mp A_{\CP+})-R_{\CP-}(1\mp A_{\CP-})}{4}\,,\\
&&r^2=x_\pm^2+y_\pm^2=\frac{R_{\CP+}+R_{\CP-}-2}{2}\,,
\end{eqnarray}
where $x_\pm=r\cos(\delta\pm\gamma)$ and
$y_\pm=r\sin(\delta\pm\gamma)$ are the same $\CP$ parameters as were 
measured by the \babar\ Collaboration with $B^-{\to} DK^-, D{\to}
  K^0_S\pi^-\pi^+$ decays~\cite{babar_dalitz}. This
choice allows the results of the two measurements to be expressed in a
consistent manner.

The measurements use a sample of 232 million
\FourS\ decays into $B\overline{B}$ pairs, corresponding to an
integrated luminosity of 211~\invfb, collected with the 
\babar\ detector at the \pep2\ asymmetric-energy $B$ factory. 
Since the \babar\ detector is described in detail elsewhere~\cite{detector},  
only the components that are crucial to this analysis are
summarized here. 
Charged-particle tracking is provided by a five-layer silicon
vertex tracker (SVT) and a 40-layer drift chamber (DCH). 
For charged-particle identification, ionization energy loss in
the DCH and SVT, and Cherenkov radiation detected in a ring-imaging
device (DIRC) are used.
Photons are identified by the electromagnetic calorimeter
(EMC), consisting of 6580 thallium-doped CsI crystals. 
These systems are mounted inside a 1.5-T solenoidal
superconducting magnet. 
We use the GEANT~\cite{geant} software to simulate interactions of
particles traversing the detector, taking into account the varying
accelerator and detector conditions. 

We reconstruct \btodh\ decays, where the prompt track $h^-$ is a kaon
or a pion. 
$D^0$ candidates are
reconstructed in the \CP-even eigenstates $\pi^-\pi^+$ and $K^-K^+$
($D^0_{\CP+}$), in the \CP-odd eigenstates $K^0_S\pi^0$, $K^0_S\phi$
and $K^0_S\omega$ ($D^0_{\CP-}$), and 
in the non-\CP, flavor eigenstate $K^-\pi^+$. $\phi$ candidates are
reconstructed in the $K^-K^+$ channel and $\omega$ candidates 
in the $\pi^-\pi^+\pi^0$ channel.
We optimize our event selection to minimize the statistical error on
the $B^-{\to} D^0_{(\CP)}K^-$ signal yield, determined for each $\Dz$
decay channel using simulated signal and background events.

The prompt particle $h$ is required to have a momentum greater than 1.4 \gevc
and the number of photons associated to its Cherenkov ring is required to be greater than four to 
improve the quality of the reconstruction. We reject a candidate track
if its Cherenkov angle does not agree within four standard deviations
($\sigma$) 
with either the pion or kaon hypothesis,
or if it is identified as an electron by the DCH and the EMC.
Particle identification (PID) information from the drift chamber and, when
available, from the DIRC, must be consistent with the kaon
hypothesis for the $K$ meson candidate in $\Dz{\to} \Km\pip$, $\Dz{\to}
\Km\Kp$, and $\phi{\to}\Km\Kp$ decays and with the pion 
hypothesis for the $\pi^\pm$ meson candidates in $D^0{\to}\pi^-\pi^+$
and $\omega{\to}\pip\pim\piz$ decays.

Neutral pions are reconstructed by combining pairs of photon candidates with
energy deposits larger than 70~\mev that are not matched to charged
tracks. The $\gamma\gamma$ invariant mass is required to be in the
range 115--150 \mevcc\ and the total \piz energy must be greater than
200 \mev.
When $\pi^0$'s are combined with other particles to form composite
particles, the mass is constrained to the nominal mass~\cite{PDG2004}.

Neutral kaons are reconstructed from pairs of oppositely charged
tracks with invariant mass within 7.8~\mevcc ($\sim 3\sigma$) of the
nominal $K^0$ mass. We also require that the ratio between the flight
length in the plane transverse to the beam direction and its error
be greater than 2. 
The $\phi$ mesons are reconstructed from two oppositely charged kaons
with invariant mass in the range
$1.008<M(K^+K^-)<1.032$~\gevcc. We also require 
$|\cos\theta_{\rm hel}(\phi)|>0.4$, where $\theta_{\rm hel}(\phi)$ is the
angle between the flight direction of one of the $\phi$ daughters and
the $D^0$ flight direction, in the $\phi$ rest frame.
The $\omega$ mesons are reconstructed from $\pi^+\pi^-\pi^0$
combinations with invariant mass in the range
$0.763<M(\pi^+\pi^-\pi^0)<0.799$~\gevcc.
We define $\theta_N$ as the angle between the normal to the $\omega$
decay plane and the $D^0$ momentum in the $\omega$ rest frame, and
$\theta_{\pi\pi}$ as the angle between the flight direction of one of
the three pions in the $\omega$ rest frame and the flight direction of
one of the other two pions in their center-of-mass (CM) frame. 
The quantities $\cos\theta_N$ and $\cos\theta_{\pi\pi}$ follow
$\cos^2\theta_N$ and $\sin^2\theta_{\pi\pi}$ distributions for the signal
and are almost flat for wrongly reconstructed or false $\omega$ candidates.
We require the product $\cos^2\theta_N\sin^2\theta_{\pi\pi}>0.08 $. 
The invariant mass of a \Dz\ candidate, $M(D^0)$, must be
within 2.5$\sigma$ of the mean fitted mass, with resolution $\sigma$
ranging from 4 to 20\mevcc depending on the $D^0$ decay mode.
For \dotopp, the invariant mass of the
$(h^-\pi^+)$ system, where $\pi^+$ is the pion from $D^0$, and $h^-$ is
the prompt track from $B^-$ taken with the kaon mass hypothesis, must
be greater than $1.9\ \gevcc$ to reject background from
$B^-{\to}D^0\pi^-, D^0{\to}K^-\pi^+$ and $B^-{\to}K^{*0}\pi^-,
K^{*0}{\to}K^-\pi^+$ decays.
When reconstructing $B$ mesons, for all $D^0$ decay channels the $D^0$
candidate invariant mass is constrained to the nominal $D^0$
mass~\cite{PDG2004}.

We reconstruct $B$ meson candidates by combining a \Dz\ candidate
with a track $h$. For the $D^0{\to}K^-\pi^+$ mode, the charge of the
track $h$ must match that of the kaon from the $D^0$ meson decay.
We select $B$ meson candidates using the beam-energy-substituted mass 
$\mes = \sqrt{(E_0^{*2}/2 + \mathbf{p}_0\cdot\mathbf{p}_B)^2/E_0^2-p_B^2}$
and the energy difference $\Delta E=E^*_B-E_0^*/2$, 
where the subscripts $0$ and $B$ refer to the initial $e^+e^-$ system and the 
$B$ candidate respectively, and the asterisk denotes the
CM ($\Upsilon(4S)$) frame.
The \mes\ distributions for \btodh\ signals are Gaussian functions
centered at the $B$ mass with a resolution of $2.6 \mevcc$, which do
not depend on the decay mode or on the nature of the prompt track.
In contrast, the \DeltaE\ distributions depend on the mass assigned to the
prompt track and on the \Dz\ momentum resolution. 
We evaluate $\Delta E$ with the kaon mass hypothesis 
so that the distributions are Gaussian and centered near zero
for \btodk\ events and shifted by approximately $50 \mev$ for \btodp\ events.
The \btodk\ \DeltaE\ resolution is about $17\mev$ for all the \Dz\
decay modes.
All $B$ candidates are selected 
with \mes within 3$\sigma$ of the peak value and with \DeltaE in the 
range $-0.16<\Delta E<0.23\gev$. 

To reduce background from continuum production of light quarks, we
construct a linear Fisher discriminant~\cite{fisher} based on the
following quantities:
(i) $L_0=\sum_i p_i$ and $L_2=\sum_i p_i\cos^2\theta_i$, evaluated in
the CM frame, where $p_i$ is the momentum, and $\theta_i$ is the 
angle with respect to the thrust axis of the $B$ candidate of charged
tracks and neutral clusters not used to reconstruct the $B$;
(ii) $|\cos\theta_T|$, where $\theta_T$ is the angle between 
the thrust axes of the $B$ candidate and of the remaining tracks 
and clusters, evaluated in the CM frame; (iii) $|\cos\theta_B|$, where
$\theta_B$ is the polar angle of the $B$ candidate in the CM frame.

For events with multiple \btodh candidates (1-7\% of the selected
events, depending on the \Dz decay mode), we choose that with the
smallest $\chi^2$ formed from the differences of the measured and true
masses of the candidate $B$, $D^0$, $\pi^0$ (only for $\Dz{\to}
K^0_S\pi^0, K^0_S\omega$), $\phi$ ($\Dz{\to} K^0_S\phi$), and $\omega$
($\Dz{\to}K^0_S\omega$), compared to the appropriate reconstructed
mass resolutions.
The total reconstruction efficiencies, based on simulated 
signal events, are 39\% ($K^-\pi^+$), 31\% ($K^-K^+$), 30\% ($\pi^-\pi^+$),
17\% ($K^0_S\pi^0$), 20\% ($K^0_S\phi$), and 7\% ($K^0_S\omega$).  

The main contributions to the background from \BB events come
from the processes $B{\to}D^{*}h$ ($h=\pi,K$), $B^-{\to}D^0\rho^-$,
mis-reconstructed \btodh, and from charmless $B$ decays to
the same final state as the signal: for instance, the process
$B^-{\to}K^-K^+K^-$  is a background for $B^-{\to}D^0K^-, D^0{\to}K^-K^+$.
These charmless backgrounds have similar \DeltaE\ and \mes
distribution as the $D^0 K^-$ signal and we call them ``peaking
\BB backgrounds''.

For each \Dz\ decay mode an extended unbinned maximum likelihood fit
to the selected data events determines yields for two signal channels,
\btodp\ and \btodk, 
and four kinds of backgrounds: candidates selected either from
continuum or from \BB\ events, in which the prompt track is either
a pion or a kaon. 
The fit uses as input \DeltaE\ and a particle identification 
probability for the prompt track based
on the Cherenkov angle $\theta_C$, the momentum $p$, and
the polar angle $\theta$ of the track.

The extended likelihood function $\cal L$ is defined as
\begin{equation}
{\cal L}= \exp\left(-\sum_{i=1}^6 n_i\right)\, \prod_{j=1}^N
\left[\sum_{i=1}^6 n_i {\cal P}_i\left(\vec{x}_j;
\vec{\alpha}_i\right) \right]\,,
\end{equation}
where $N$ is the total number of observed events and $n_i$ is the
yield of the $i^{th}$ event category. 
The six functions ${\cal P}_i(\vec{x}_j;\vec{\alpha}_i)$ are the
probability density functions (PDFs) for the variables
$\vec{x}_j$, given the set of parameters $\vec{\alpha}_i$. They are
evaluated as a product
$\mathcal{P}_i=\mathcal{P}_{1i}(\DeltaE)\times\mathcal{P}_{2i}({\theta_C})$.

The \DeltaE\ distribution for \btodk\ signal events is parametrized
with a Gaussian function.
The \DeltaE\ distribution for \btodp\ is parametrized with the same
Gaussian function used for \btodk\ with an additional shift, computed event
by event as a function of the prompt track momentum, arising from the
wrong mass assignment to the prompt track. The offset and width of
the Gaussian functions are determined from data together with the yields.
 
The \DeltaE\ distribution for the continuum background is parametrized with
a linear function whose slope is determined from off-resonance data.
The \DeltaE\ distribution for the non-peaking \BB background is empirically
parametrized with the sum of a Gaussian function and an exponential
function when the prompt track is a pion, and with an exponential
function when the prompt track is a kaon. The parameters are
determined from simulated events. The \DeltaE\ distribution for the
peaking charmless \BB background is parametrized with the same
Gaussian function used for the \btodk signal.
The yield of the \BB peaking background is estimated
from the sidebands of the $D^0$ invariant mass distribution and
fixed in the fit.

The parametrization of the particle identification PDF is performed 
by fitting with two Gaussian functions the background-subtracted
distribution of the difference between the reconstructed and expected
Cherenkov angles of kaon and pion samples. The parametrization
is performed as a function of the momentum and polar angle of the
track. Pions and kaons are selected from a pure $D^{*+}\to\Dz\pip$,
$\Dz{\to}\Km\pip$ control sample.

\begin{figure}[!htb]
\begin{center}
\includegraphics[width=7.5cm,height=3.6cm]{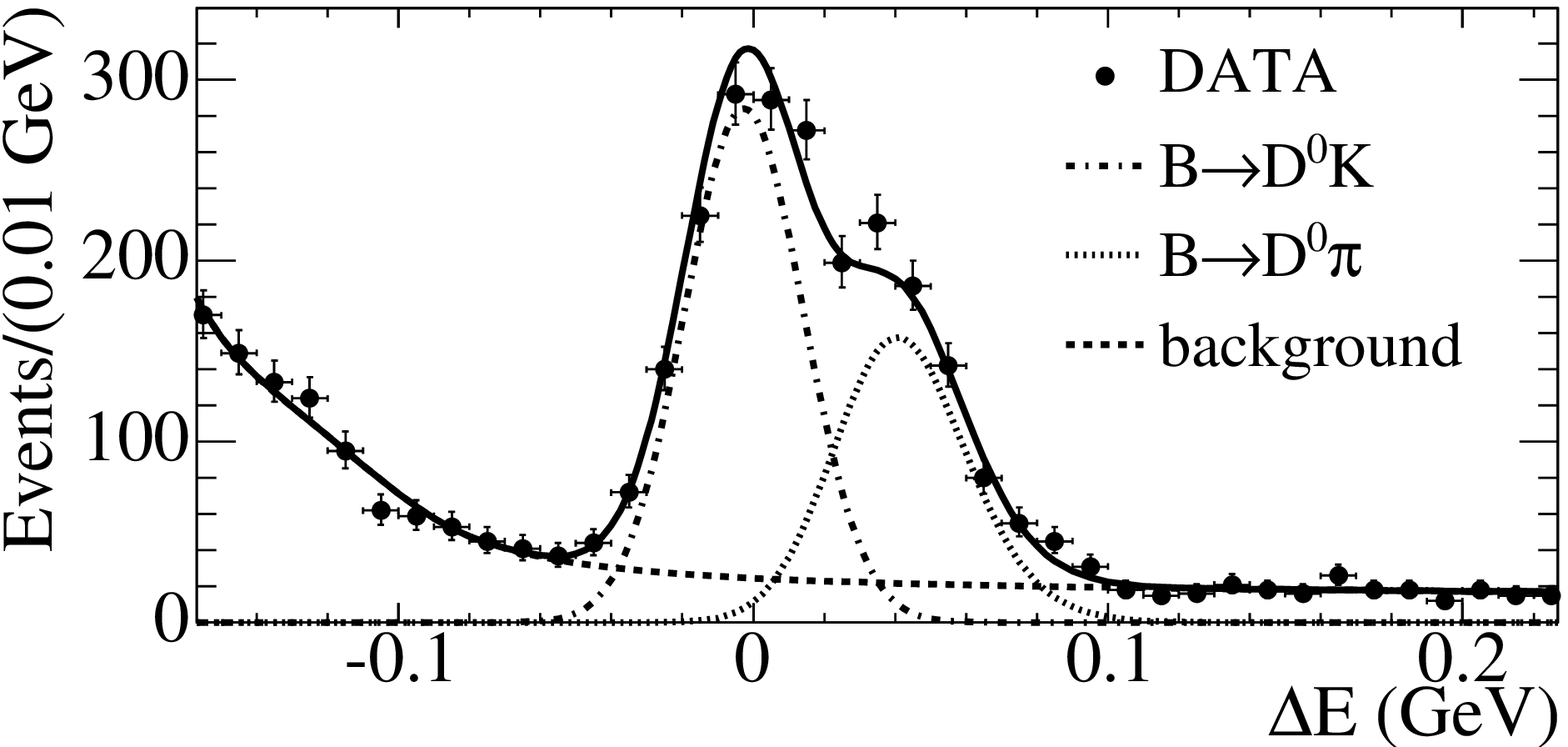}
\includegraphics[width=7.5cm,height=3.6cm]{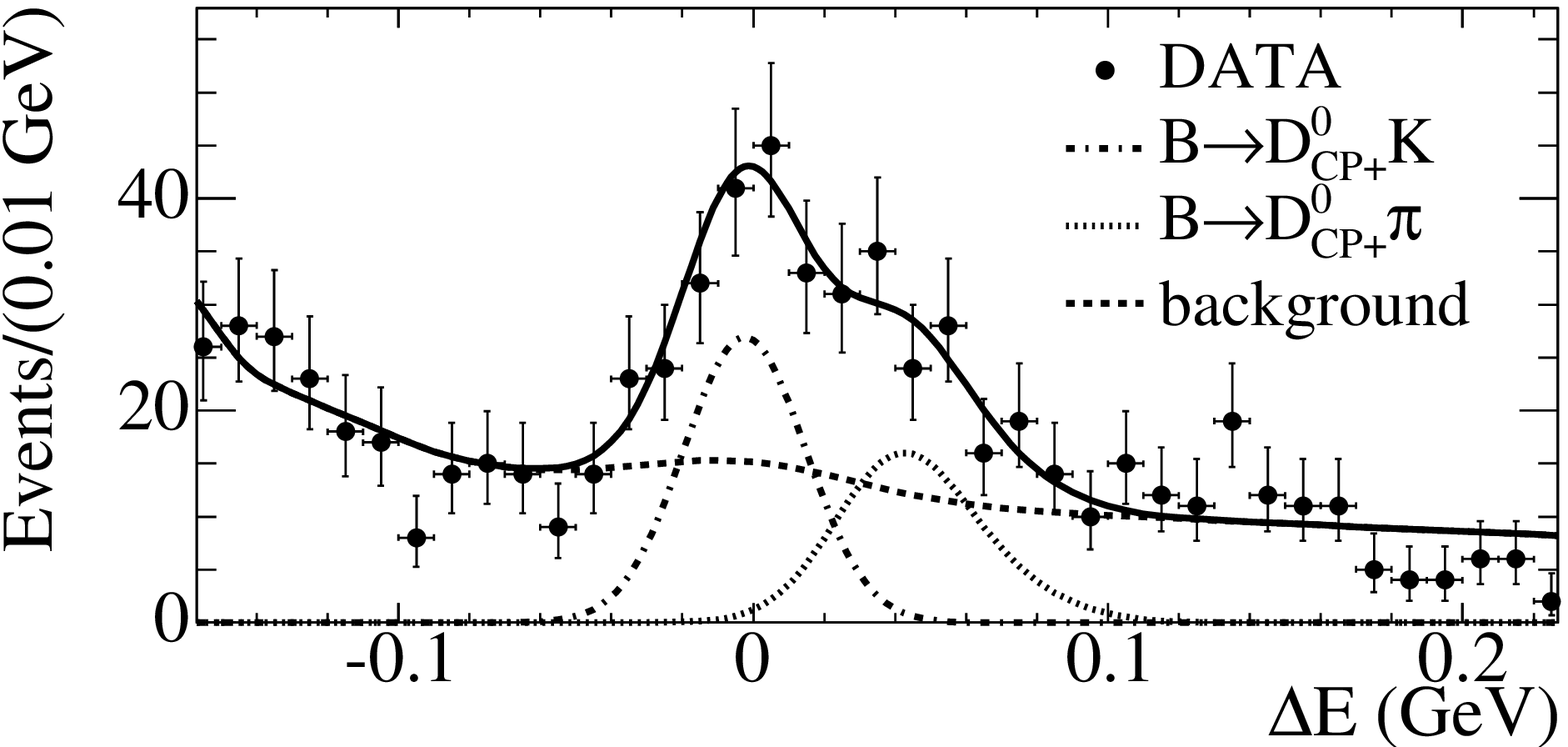}
\includegraphics[width=7.5cm,height=3.6cm]{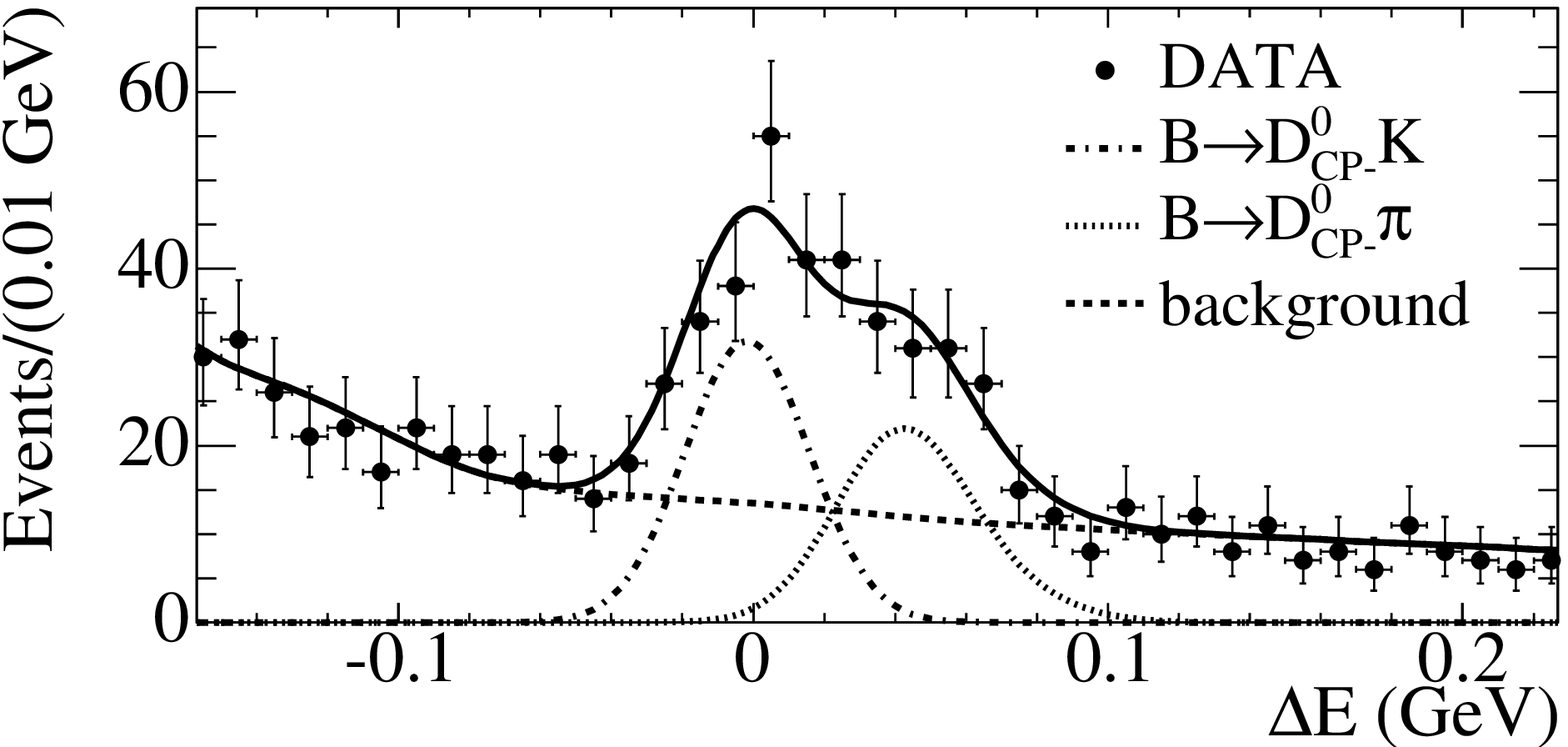}
\caption{Distributions of $\Delta E$ for events enhanced in
  $B^-{\to}D^0K^-$ signal. Top: $B^-{\to}D^0K^-, D^0{\to}K^-\pi^+$; middle:
  $B^-{\to}D^0_{\CPp}K^-$; bottom: $B^-{\to}D^0_{\CPm}K^-$. Solid curves
  represent projections of the maximum likelihood fit; dashed,
  dashed-dotted and dotted curves represent the $B{\to}D^0K$,
  $B{\to}D^0\pi$ and background contributions.} 
\label{fig:fit_kaons}
\end{center}
\end{figure}

The results of the fit are summarized in Table~\ref{tab:fitresults}.
Figure~\ref{fig:fit_kaons} shows the distributions of \DeltaE\ for the 
$K^-\pi^+$, \CPp\ and \CPm\ modes after enhancing the $B{\to}D^0K$
purity by requiring that the prompt track be consistent with the 
kaon hypothesis.
The total PDF, normalized by the fitted signal and background yields,
integrated over the Cherenkov angle variable and modified to take into
account the tighter selection criteria, is overlaid in the figure.

\begin{table}[!htb]
\caption{Yields from the maximum likelihood fit.}
\label{tab:fitresults}
\begin{center}
\begin{tabular}{lcccc}
\hline
\hline
$D^0$ mode &\  $N(D\pi^+)$ &\ $N(D\pi^-)$& $N(DK^+)$ & $N(DK^-)$\\
\hline
$K^-\pi^+$           &\  $8151\pm 95$ &\  $7899 \pm 93$ & $649\pm29$ & $611\pm28$\\
\hline
$K^-K^+$           &\  $705\pm 28$ &\  $690 \pm 28$ & $26\pm 9$ & $70\pm 10$\\
$\pi^-\pi^+$       &\  $256\pm 18$ &\  $219 \pm 17$ & $18\pm 7$ & $17\pm 7$\\
\hline
$K^0_S\pi^0$           &\  $707\pm 29$ &\  $677 \pm 29$ & $39\pm 9$ & $42\pm 9$\\
$K^0_S\phi$            &\  $176\pm 14$ &\  $157 \pm 13$ & $15\pm 5$ & $13\pm 4$\\
$K^0_S\omega$          &\  $235\pm 17$ &\  $230 \pm 17$ & $25\pm 7$ & $14\pm 6$  \\
\hline
\hline
\end{tabular}
\end{center}
\end{table}

The ratios $R_{\CP\pm}$ are computed for the five $\CP$ modes using
the relations in Eqs.~(\ref{eq:r}) and~(\ref{eq:rpm}).
A number of systematic uncertainties, such as the uncertainty
from the tracking efficiency and the uncertainty on the $D^0$
decay branching fractions,
cancel out in the measurement of the double ratio.
The relations $R_{\CP\pm}=R_\pm/R$ hold under the assumption that the
magnitude $r_\pi$ of the ratio of the amplitudes of the $B^-{\to}
\Dzb\pi^-$ and $B^-{\to} D^0\pi^-$ processes can be
neglected~\cite{gronau2003} ($r_\pi \sim r
\frac{\lambda^2}{1-\lambda^2} \lesssim 0.012$, where 
$\lambda \approx 0.22$~\cite{PDG2004} 
is the sine of the Cabibbo
angle). This assumption is considered further when we discuss the
systematic uncertainties.
The quantities $R_{\pm}/R$ are computed
from the ratios of the $B{\to}DK$ and $B{\to}D\pi$ yields in
Table~\ref{tab:fitresults}, scaled by correction factors taking into 
account small differences in the selection efficiency between $B{\to}DK$ and
$B{\to}D\pi$.
These correction
factors are evaluated from simulated events and range between
$0.982\pm 0.018$ and $1.020\pm 0.031$ depending on the \Dz decay mode.
The results for the \CP-even and \CP-odd combinations are listed in 
Table~\ref{tab:final_ratio}.
 
The partial-rate charge asymmetries $A_{\CP\pm}$ are calculated from
the measured yields of positive and negative $B{\to}DK$ decays in
Table~\ref{tab:fitresults}. 
The results for the \CP-even and \CP-odd combinations are reported in
Table~\ref{tab:final_ratio}.

\begin{table}[!htb]
\caption{Measured ratios $R_{\CP\pm}$ and $A_{\CP\pm}$
for $\CP$-even and $\CP$-odd $D$ decay modes.
The first error is statistical, the second is
systematic. $R_{\CP-}$ and $A_{\CP-}$ are corrected for the \CP-even dilution
described in the text.}
\label{tab:final_ratio}
\begin{center}
\begin{tabular}{lcc}
\hline
\hline
$D^0$ mode &\ \ \ \ \ \ $R_{\CP}$ &\ \ \ \ \ \ $A_{\CP}$\\
\hline
$\CP+$ &\ \ \ \ \ \  $0.90\pm 0.12 \pm 0.04$&\ \ \ \ \ \ \ \ $0.35\pm 0.13 \pm 0.04$ \\
$\CP-$ &\ \ \ \ \ \  $0.86\pm 0.10 \pm 0.05$&\ \ \ \ \ \ $-0.06\pm 0.13 \pm 0.04$ \\
\hline
\hline
\end{tabular}
\end{center}
\end{table}

In the case of $D^0{\to}\KS \phi$, $\phi{\to}K^+K^-$, and
$D^0{\to}\KS \omega$, $\omega{\to}\pi^+\pi^-\pi^0$, the values of
$R_{\CP-}$ and $A_{\CP-}$ 
quoted in Table~\ref{tab:final_ratio} are obtained after correcting the
measured values to take into account the dilution from
a \CP-even background arising from $B^-{\to}\Dz h^-$,
$\Dz{\to}K^0_S(K^-K^+)_{\textrm{non}-\phi}$ and 
$\Dz{\to}K^0_S(\pi^-\pi^+\pi^0)_{\textrm{non}-\omega}$ decays. 
For the $K^0_S \phi$ channel we exploit the investigation performed
by \babar\ of the $\Dz{\to}\KS K^+ K^-$ Dalitz
plot~\cite{babar_dztokskk} to estimate the level of the \CP-even
background ($0.160\pm0.006$ relative to the $\KS\phi$ 
signal) and the corresponding $R_{\CP-}$ and $A_{\CP-}$ dilution. For
the $K^0_S \omega$ channel there is little information on this
background. We estimate the amount of $\Dz{\to}
K^0_S(\pi^+\pi^-\pi^0)_{\textrm{non}-\omega}$ background ($0.25\pm 0.05$
relative to the $\KS \omega$ signal) from the $\cos\theta_N$
distribution of $B^-{\to}D^0\pi^-$, $\Dz{\to}\KS\pi^+\pi^-\pi^0$
candidates, and assume the \CP-even content of this background
to be $(50\pm 29)\%$.

Systematic uncertainties in the ratios $R_{\CP\pm}$ and in the \CP 
asymmetries $A_{\CP\pm}$ are listed in
Table~\ref{tab:syst}. They arise both from the uncertainties on the
signal yields, extracted through the maximum likelihood fit,
and from the assumptions used to compute $R_{\CP\pm}$ and
$A_{\CP\pm}$. The correlations between the different sources of
systematic errors, when non-negligible, are considered when combining
the two \CP-even or the three \CP-odd modes.

\begin{table}[!htb]
\caption{Systematic uncertainties on the observables $R_{\CP\pm}$ and
  $A_{\CP\pm}$ after combination of the two \CP-even and the three
  \CP-odd \Dz decay modes.}
\label{tab:syst}
\begin{center}
\begin{tabular}{lcccc}
\hline
\hline
Source &  $\Delta R_{\CP+}$ & $\Delta R_{\CP-}$ & $\Delta A_{\CP+}$ &
$\Delta A_{\CP-}$ \\
       &  $(\%)$            & $(\%)$     & $(\%)$     & $(\%)$     \\ 
\hline
bkg. $\DeltaE$ PDF   & 1.3 & 1.1 & 1.1 & 0.4 \\
PID PDF              & 0.1 & 0.1 & 0.2 & 0.2 \\
peaking bkg. yields  & 3.0 & 4.2 & 2.6 & 2.2 \\
opposite-\CP bkg.    & -   & 1.3 & -   & 1.0 \\
detector charge asym.& -   & -   & 2.7 & 2.7 \\
$\varepsilon^{K/\pi}_{\pm}/\varepsilon^{K/\pi}$  & 1.0 & 1.1 & - & - \\
$r_\pi$ & 2.2 & 2.1 & - & - \\
\hline
{\bf Total} & 4.1 & 5.1 & 3.9 & 3.7 \\
\hline
\hline
\end{tabular}
\end{center}
\end{table}

The uncertainties on the fitted signal yields are due to the
imperfect knowledge of the $\DeltaE$ and PID PDFs and of the 
peaking background yields, and are evaluated by varying the
parameters of the PDFs and the peaking background yields by $\pm1\sigma$ 
and taking the difference in the signal yields. 
The uncertainties in the branching fractions used in the simulation of
the $B$ decays that contribute to the \BB\ background are also taken
into account. The yields of the \BB\ and continuum backgrounds
found in data are consistent with what is expected from the simulation.
In the $K^0_S\phi$ and $K^0_S\omega$ channels we also take into
account the uncertainties in the dilution factors due to the imperfect
knowledge of the levels of the \CP-even backgrounds from
$B^-{\to}D^0K^-$, $\Dz{\to}K^0_S(K^-K^+)_{\textrm{non}-\phi}$ and 
$\Dz{\to}K^0_S(\pi^-\pi^+\pi^0)_{\textrm{non}-\omega}$ decays. 

A possible bias in the measured $A_{\CP\pm}$ may come
from an intrinsic detector charge asymmetry due to asymmetries in
acceptance or tracking and particle identification efficiencies.
An upper limit on this bias has been obtained from the measured
asymmetries in the processes $B^-{\to}D^0h^-, D^0{\to}K^-\pi^+$ and 
$B^-{\to}D^0_{\CP\pm}\pi^-$, where \CP\ violation is expected to be
negligible. From the average asymmetry, ($-1.8\pm 0.9)\%$, we obtain the
limit $\pm 2.7\%$ for the bias. This has been added in quadrature to
the total systematic uncertainty on the \CP\ asymmetry.

For the branching fraction ratios $R_{\CP\pm}$ two additional sources
of uncertainty are the correction factors used to scale the yield
ratios, and the assumption that $R_{\CP\pm} = R_{\pm}/R$. The scaling
factor, estimated from simulated events, is a double ratio of efficiencies,
$\varepsilon^{K/\pi}_{\pm}/\varepsilon^{K/\pi}$, where
$\varepsilon^{K/\pi}_{(\pm)}$ denotes the ratio between the selection
efficiencies of $B{\to}D^0_{(\CP\pm)}K$ and
$B{\to}D^0_{(\CP\pm)}\pi$. In the double ratio the systematic
uncertainties 
arising from possible discrepancies between data and
simulation are negligible, and only the contribution from the limited
statistics of the simulated samples remains. The assumption
$R_{\CP\pm} = R_{\pm}/R$ introduces a relative uncertainty $\pm 2 r_\pi
\cos\delta_\pi \cos\gamma$ on $R_{\CP\pm}$, where $\delta_\pi$ is the
relative strong phase between the amplitudes $A(B^-{\to}\Dzb \pi^-)$
and $A(B^-{\to}\Dz \pi^-)$. Since $|\cos\delta_\pi \cos\gamma|\le 1$ and
$r_\pi \lesssim 0.012$, we assign a relative uncertainty $\pm 2.4\%$ to
$R_{\CP\pm}$, which is completely anti-correlated between $R_{\CP+}$
and $R_{\CP-}$.

We quote the measurements in terms of $x_\pm$ and $r^2$,
\begin{eqnarray}
&&x_+=-0.082\pm 0.053\stat\pm0.018\syst\,,\\
&&x_-=+0.102\pm 0.062\stat\pm0.022\syst\,,\\
&&r^2=-0.12\pm0.08\stat\pm 0.03\syst.
\end{eqnarray}
The measured values of $x_\pm$ are consistent with those found, on a
slightly smaller data sample, with the $B^-{\to}DK^-$,
$D{\to}K^0_S\pi^-\pi^+$ decays, and the precision is
comparable~\cite{babar_dalitz}. The measured value of $r^2$ is
consistent with the upper limits on $r$ from \babar\ and
Belle~\cite{babar_ads}. 

In conclusion, we have reconstructed \btodk\ decays with
$D^0$ mesons decaying to non-\CP, \CP-even and \CP-odd eigenstates.
We have improved the previous measurements of $R_{\CP\pm}$ and
$A_{\CP\pm}$~\cite{babarbtdk,bellebtdk}, and we have also expressed the
results in terms of the 
same $x_\pm$ parameters as were measured with $B^-{\to}DK^-$,
$D{\to}K^0_S\pi^-\pi^+$ 
through a Dalitz plot analysis of the $D$ final
state~\cite{babar_dalitz}, with a comparable precision.
These measurements, combined with the existing measurements of the
$B{\to}DK$ decays, will improve the knowledge of the angle $\gamma$ and the
parameter $r$.

We are grateful for the excellent luminosity and machine conditions
provided by our \pep2\ colleagues, 
and for the substantial dedicated effort from
the computing organizations that support \babar.
The collaborating institutions wish to thank 
SLAC for its support and kind hospitality. 
This work is supported by DOE and NSF (USA),
NSERC (Canada), IHEP (China), CEA and CNRS-IN2P3
(France), BMBF and DFG (Germany), INFN (Italy), FOM (the Netherlands),
NFR (Norway), MIST (Russia), and PPARC (United Kingdom). 
Individuals have received support from the 
A.~P.~Sloan Foundation, Research Corporation,
and Alexander von Humboldt Foundation.

\end{document}

%% file: authors_nov2005.tex
%
\author{B.~Aubert}
\author{R.~Barate}
\author{D.~Boutigny}
\author{F.~Couderc}
\author{Y.~Karyotakis}
\author{J.~P.~Lees}
\author{V.~Poireau}
\author{V.~Tisserand}
\author{A.~Zghiche}
\affiliation{Laboratoire de Physique des Particules, F-74941 Annecy-le-Vieux, France }
\author{E.~Grauges}
\affiliation{IFAE, Universitat Autonoma de Barcelona, E-08193 Bellaterra, Barcelona, Spain }
% ...
\collaboration{The \babar\ Collaboration}
\noaffiliation

%% file: paper.bbl
\begin{thebibliography}{99}
\bibitem{chargeconj}
Reference to the charge-conjugate state is implied
here and throughout the text unless otherwise stated.
\end{thebibliography}
